\documentclass{PoS}

\title{KKLT and the Swampland Conjectures\footnote{Report Numbers: MPP-2020-56, MITP/20-024}}

\ShortTitle{KKLT and the Swampland Conjectures}

\author{\speaker{Ralph Blumenhagen} \\
        Max-Planck-Institut f\"ur Physik, F\"ohringer Ring 6,  80805 M\"unchen, Germany \\
        E-mail: \email{blumenha@mpp.mpg.de}}

\author{Max Brinkmann \\
        Max-Planck-Institut f\"ur Physik, F\"ohringer Ring 6,  80805 M\"unchen, Germany \\
        E-mail: \email{mbrinkm@mpp.mpg.de}}
        
\author{Daniel Kl\"awer\\
     PRISMA Cluster of Excellence and Mainz Institute for Theoretical Physics, Johannes Gutenberg-Universit\"at, 55099 Mainz, Germany\\
        E-mail: \email{klaewer@uni-mainz.de}}

\author{Andriana Makridou \\
        Max-Planck-Institut f\"ur Physik, F\"ohringer Ring 6,  80805 M\"unchen, Germany \\
         E-mail: \email{amakrido@mpp.mpg.de}}
 
\author{Lorenz Schlechter \\
        Max-Planck-Institut f\"ur Physik, F\"ohringer Ring 6,  80805 M\"unchen, Germany \\
        E-mail: \email{lschlech@mpp.mpg.de}}

\abstract{Recently, various swampland conjectures have been proposed
  that every UV complete effective field theory should satisfy.
In particular, these deal with the properties of AdS and dS solutions
arising in theories of quantum gravity.
Mainly summarizing the results of our two previous articles 
\cite{Blumenhagen:2019qcg,Blumenhagen:2019vgj},
we confront the string theory based KKLT scenario with these conjectures and argue that
if quantum vacua as in the KKLT construction indeed exist, 
some of the conjectures receive some $\log$-corrections.
Furthermore we point out some new aspects not contained in the
aforementioned two papers.
}

\FullConference{Corfu Summer Institute 2019 "School and Workshops on Elementary Particle Physics and Gravity" (CORFU2019)\\
	31 August - 25 September 2019\\
	Corfu, Greece}

\usepackage{amsmath}
\usepackage{amsfonts}
\usepackage{amssymb}
\usepackage{graphicx}
\usepackage{tikz} 
\usetikzlibrary{decorations.pathreplacing,calc}
\usepackage[utf8]{inputenc}
\usepackage{cite}

\newcommand{\eq}[1]{\begin{equation}
	\begin{split} #1 \end{split}
	\end{equation}}

\newcommand{\ov}{\overline}

\allowdisplaybreaks[2]
\numberwithin{equation}{section}

\begin{document}

\section{Introduction}
\label{sec_intro}

During the last years, the idea of a string or quantum gravity
swampland \cite{Vafa:2005ui,Ooguri:2006in} was further developed into various directions and in the moment can be considered the
hot topic in the field of string phenomenology/cosmology.
As opposed to the string theory landscape, the swampland 
contains those low-energy effective theories that cannot be 
UV completed to a consistent theory of quantum gravity.
The aim of the swampland program is  to extract a set of
relatively simple  features 
that  low-energy effective field theories should satisfy in order to admit such an embedding into a theory of quantum gravity
(see~\cite{Brennan:2017rbf,Palti:2019pca} for recent reviews). This makes string theory much more predictive than previously expected from the picture of a vast string landscape.

Several such swampland conjectures have been proposed. In this article
we focus on a number of them that mainly deal with the properties 
of anti-de Sitter (AdS) and de Sitter solutions in string
theory\footnote{Recently, there was also a conjecture concerning
  Minkowski minima in string theory\cite{Palti:2020qlc}. This  says that in 4D theories
  with genuine  ${\cal N}=1$  supersymmetry there are no  exact Minkowski minima.}.
First, these are the AdS/moduli scale separation conjecture (AM-SSC)
\cite{Gautason:2018gln} and the AdS distance conjecture (ADC) 
\cite{Lust:2019zwm} (see also \cite{Alday:2019qrf,Kehagias:2019akr}). 
These conjectures say that the mass of certain (towers of) modes 
cannot be parametrically separated from the AdS radius.
The strong version of the ADC is reminiscent of observations 
made earlier in \cite{Gautason:2015tig}.
Second there are conjectures that are related to dS solutions, namely 
the dS swampland conjecture
\cite{Obied:2018sgi,Ooguri:2018wrx,Garg:2018reu,Andriot:2018wzk}
and its quantum version called the trans-Planckian censorship
conjecture (TCC) \cite{Bedroya:2019snp}.
This quantum generalization
admits meta-stable dS extrema as long as their life-time is
sufficiently small. 
In addition we also discuss the  emergence proposal 
\cite{Heidenreich:2017sim,Grimm:2018ohb,Heidenreich:2018kpg} of infinite
distances in field space that is closely related to the swampland
distance conjecture \cite{Ooguri:2006in,Klaewer:2016kiy}.

These conjectures have been motivated and tested in the framework of  tree-level 
constructions of AdS and dS vacua.
However, it is well known that there exist also constructions of  such
vacua that utilize
not only tree-level ingredients, but also quantum, in particular
non-perturbative effects. The most famous examples are the KKLT \cite{Kachru:2003aw} 
and the large volume scenario (LVS) \cite{Balasubramanian:2005zx}. 
In both  cases, AdS minima are found in the effective 4D potential
and subsequently uplifted to dS.
This article is essentially a brief review of recent work on relating the
KKLT construction to the aforementioned swampland conjectures.

In section \ref{sec_one} we briefly review the relevant AdS and dS
swampland conjectures, followed in section \ref{sec_flux} by 
the presentation of two prototypical  AdS flux vacua arising in the class
of type IIA orientifolds. 
Section \ref{sec_KKLT} provides a description of the KKLT construction
in the framework of an effective theory which takes into account that
for a successful uplift,  a strongly warped throat geometry is required.
In fact, starting with \cite{Bena:2018fqc} there were a couple of recent papers
that deal with such a warped KKLT model both from a stringy 10D
perspective \cite{Blumenhagen:2019qcg,Bena:2019sxm,Dudas:2019pls}  and from an effective 5D point of view \cite{Randall:2019ent}.
In this proceedings article  we will mostly focus on our own  work  \cite{Blumenhagen:2019qcg}, where 
in particular the existence of an exponentially light tower of KK
modes was established. This also fits nicely  into the scheme of the emergence
proposal. In section \ref{sec_quantumLog} we discuss the relation
of the KKLT construction to the AdS and dS swampland conjectures.
Taken at face  value, the KKLT  model does not satisfy the two AdS
conjectures but would do so once one introduces quantum
$\log$-corrections to the swampland conjectures \cite{Blumenhagen:2019vgj}. 
The quantum contributions found are reminiscent of a similar
correction in the TCC. We also include some new aspects into this
review article such as a closer investigation of the dS saddle point
that the KKLT model develops once one includes $\alpha'$ corrections
to the K\"ahler potential.

\section{Swampland conjectures}
\label{sec_one}

In this section we briefly review some of the swampland conjectures
that have been proposed. First, we consider swampland conjectures
related to AdS and dS solutions. Second, we review the emergence
proposal for (infinite) distances in field space.

\subsection{AdS and dS swampland conjectures}

While dS vacua are notoriously difficult 
to get in concrete string theory constructions, AdS solutions
already exist at the classical level. The most famous example
is certainly the AdS$_5\times S^5$ solution of the type IIB
superstring. We distinguish two conjectures that deal with 
the property of these AdS spaces. The first one gives
a relation between the lightest (non-zero) moduli mass 
and the scale of the AdS space \cite{Gautason:2018gln}. 

\vspace{0.3cm}
\noindent
\underbar{ AdS/moduli scale separation conjecture} (AM-SSC): 

\noindent
{\it In an AdS minimum one cannot separate the size of the AdS space 
and the mass of its lightest mode. Quantitatively, they satisfy the relation
\eq{
\label{mscalesep}
                           m_{\rm mod}\, R_{\rm AdS} \le c 
}
where $R^2_{\rm AdS}\sim |\Lambda|^{-1}$ and $c$ denotes an order one parameter.
}
\vspace{0.3cm}

\noindent
To our knowledge, this conjecture is satisfied in all known tree-level
AdS flux vacua. We will see later that for the  KKLT AdS minimum, 
relation \eqref{mscalesep} receives  $\log$-corrections.
The second swampland conjecture is reminiscent of the swampland
distance conjecture and relates the AdS scale to the mass scale of
a tower of light states~\cite{Lust:2019zwm}.

\vspace{0.3cm}

\noindent
\underbar{ AdS distance conjecture} (ADC):  

\noindent
{\it For an  AdS vacuum the limit $\Lambda\to 0$ is at infinite distance
in field space and there is a tower of light states with
\eq{
\label{AdSdis}
                    m_{\rm tower}= c_{\rm AdS} \, |\Lambda|^{\alpha} 
}
for  $\alpha>0$. }

\vspace{0.3cm}
\noindent
In the original paper \cite{Lust:2019zwm}, a stronger version was proposed saying
that  for supersymmetric AdS minima one always finds $\alpha=1/2$, effectively 
forbidding any scale separation of internal and AdS space. However,
subsequent work \cite{Grimm:2019ixq,Junghans:2020acz,Buratti:2020kda,Marchesano:2020qvg} supports  
the original claim of  DGKT \cite{DeWolfe:2005uu} that 
supersymmetric type IIA flux  compactifications  provide a class of 
scale separated AdS vacua. As will be recalled in section
\ref{sec_flux},
these models feature a  four-form flux that is unbounded 
by any tadpole cancellation condition and leads to a KK scale 
$m_{\rm KK}\sim |\Lambda|^{7/18}$, hence violating the strong AdS
distance conjecture. It was suggested in \cite{Blumenhagen:2019vgj} that irrespective of
supersymmetry the strong ADC holds for models where curvature terms are relevant for
moduli stabilization.  In the framework of  flux compactifications this is described
by turning on so-called geometric fluxes. The $AdS_5\times S^5$
background is of this type when described by an effective 5D
theory. Another  example will be presented in section \ref{sec_flux}.

The conjecture with probably the boldest cosmological implications, 
though stated as a  general bound on the scalar potential in a theory of  
quantum gravity \cite{Obied:2018sgi,Ooguri:2018wrx}, 
deals with the existence of dS vacua\footnote{Motivated by  \cite{Hertzberg:2007wc},
  initially in \cite{Obied:2018sgi}  only the first relation in \eqref{dswamp} was
  proposed. This would exclude also dS saddle points that are believed
  to exist in the string theory landscape.}.

\vspace{0.2cm}
\noindent
\underbar{Refined dS swampland conjecture}:  

\noindent
{\it  The scalar potential  satisfies either of the two properties
\eq{
\label{dswamp}
                  |\nabla V|\ge {c\over  M_{\rm pl}} \cdot V\,\qquad {\rm
                    or}\qquad {\rm min}(\nabla_i\nabla_j
                  V)\le -{c'\over M_{\rm pl}^2} \cdot V  \,,
}
forbidding de Sitter minima.}

\vspace{0.3cm}
\noindent
Note that for AdS vacua the first condition is always satisfied. The second
condition means that dS extrema are always saddle points that should
decay fast enough. Indeed, the life-time of such a dS saddle point 
has been estimated in \cite{Bedroya:2019snp} as 
\eq{
\label{lifetac}
   T\sim- {3H\over m_{\rm tach}^2},
}
where $m_{\rm tach}^2$ is defined as the smallest negative eigenvalue of the mass$^2$ operator, so we get the bound  $T\le (c'H)^{-1}$. All stringy dS saddle points that we are aware of do indeed satisfy this conjecture.
Note that for our universe this would mean that the
life-time is shorter than its present life-time. Therefore, if this
conjecture is correct, it seems that quintessence\footnote{
See \cite{Akrami:2018ylq} for a compilation of the harsh observational bounds on stringy models of quintessence.} 
would remain the only natural
candidate to explain the cosmological $\Lambda$CDM model 
\cite{Caldwell:1997ii,Agrawal:2018own}.

In looking for a conceptual explanation of the dS swampland
conjecture, a slightly weaker form was proposed~\cite{Bedroya:2019snp}.
This is the trans-Planckian censorship conjecture (TCC), which 
states that sub-Planckian fluctuations should not become classical (in an
expanding universe).
Quantitatively it says
\eq{
              \int_{t_i}^{t_f} dt\, H < \log\left( {M_{\rm pl}\over H_f}\right)\,,
}            
which implies a weaker bound on the life-time of a dS phase
\eq{
\label{TCClifetime}
                     T\le {1\over H}\log\left({M_{\rm pl}\over H}\right) \,.
}
We think it would be interesting to identify stringy dS extrema that
saturate this bound, i.e. including the $\log$-factor. In this case, the present expansion
of our universe  could also be a consequence of a hill-top phase of rolling.

The TCC implies a weaker form also for the first condition in
the dS swampland conjecture  \eqref{dswamp}.
Namely,  for a monotonically  decreasing positive potential, it  yields  a  global bound 
\eq{
\label{TCCdS}
        M_{\rm pl}\left\langle{-V'\over
            V}\right\rangle\bigg\vert_{\phi_i}^{\phi_f}>{1\over \Delta \phi}
\log\Big( {V_i\over A}\Big)  +{2\over \sqrt{(d-1)(d-2)}} 
}
for  the average of $-V'/V$ in the interval $[\phi_i,\phi_f]$. 
Here $A= M_{\rm pl}^4 (d-1)(d-2)/2$ is a constant.
In its
asymptotic limit it provides a prediction of the parameter
$c=2/\sqrt{(d-1)(d-2)}$ in \eqref{dswamp}.
Recently, the TCC was also related to the swampland distance conjecture
\cite{Brahma:2019vpl,Andriot:2020lea}.

\subsection{Emergence proposal}

The emergence proposal \cite{Heidenreich:2017sim,Grimm:2018ohb,Heidenreich:2018kpg} 
can be considered as an extension of the 
swampland distance conjecture\footnote{A review talk on the relation
of large field inflation in string theory and the swampland distance
conjecture was given at the Corfu Summer School 2017 \cite{Blumenhagen:2018hsh}.}
 \cite{Ooguri:2006in}. The latter says that
approaching points at infinite distance $\phi\to\infty$ in
the moduli space of an  effective field theory, a tower of states becomes 
exponentially light
\eq{
                    m_{n} \sim f(n)\, e^{-\lambda \phi}\,,
}
with $f(n)=n$ for KK towers, or $f(n)=\sqrt{n}$ for the denser weakly coupled string spectrum \cite{Lee:2018urn, Lee:2019xtm, Lee:2019wij, Baume:2019sry}. This leads to a breakdown of the effective action for
$\phi>\lambda^{-1}$. Moreover, this critical scale is of the order
of the Planck scale.

\vspace{0.2cm}
\noindent
\underbar{Emergence proposal}: 

\noindent
{\it The  infinite distance in the IR appears from integrating out
the tower of  light states in the UV. }

\vspace{0.3cm}
\noindent
To make this more concrete, consider 
a light field $\phi$ and a tower of massive states $h_n$ with mass $m_n=n\, \Delta m(\phi)$ 
and degeneracy   ${\rm deg}(n)=n^K$ governed by an effective action
\eq{
\label{towereffact}
           S=M_{\rm pl}^{D-2} \int d^D x \left( {1\over 2} G_{\phi\phi} \partial_\mu\phi  \partial^\mu
           \phi+   \sum_n  {1\over 2}  \partial_\mu h_n \partial^\mu
           h_n +  {1\over 2} m^2_n(\phi)  h_n^2 \right)\,
}
where $G_{\phi\phi}$ denotes the metric on field space.
Integrating out the modes that are lighter than  the species scale
\eq{
                   \Lambda_{\rm sp}={\Lambda_{\rm UV} \over {N_{\rm
                           sp}}^{\frac{1}{D-2}}} 
}
leads to a one-loop correction to the metric  $G_{\phi\phi}$ as
\eq{
                  G_{\phi\phi}^{\rm 1-loop}&\sim {1\over M_{\rm pl}^{D-2}}\sum_{n=1}^{N_{\rm sp}}
                  \Big( \partial_\phi m_n(\phi) \Big)^2
                  \sim  {\Lambda_{\rm sp}^{D+K-1} \over
          M_{\rm pl}^{D-2}} {  \big( \partial_\phi \Delta
                     m(\phi) \big)^2 \over \big( \Delta
                     m(\phi) \big)^{K+3}}\,.
}
The emergence proposal then calls for  $G_{\phi\phi}^{\rm 1-loop} \sim
G_{\phi\phi}^{\rm tree}$. In its stronger form  it even claims that the
initial tree-level metric is vanishing in the UV, and that it is entirely
emergent from integrating out massive modes.

\section{Tree-level flux vacua}
\label{sec_flux}

In this section we remind the reader of  two classes of AdS minima that
occur in tree-level flux compactifications of  type IIA orientifolds.
The first class are the so-called DGKT models
\cite{DeWolfe:2005uu,Camara:2005dc} (see also
\cite{Marchesano:2019hfb} for more recent studies) for which solely various NS-NS and
R-R $p$-form fluxes are turned on. In the second class one also allows 
geometric fluxes which can be considered as generalizations of the
$AdS_5\times S^5$ Freud-Rubin type vacua.

As we will see,  for properly deciding  whether these well-known vacua
abide by the conjectures,  we need to identify the relevant scales.
Concerning the subcase of the ADC for which the relevant tower of
light states is the Kaluza-Klein one, we critically address 
the usual estimate for the KK scale $m_{\rm KK}\sim 1/R$, 
where $R$ is the relevant length scale for 
the compactification manifold.

\subsection{Example I: Type IIA flux compactifications}

The special feature of this class of flux compactifications is that
all closed string moduli can be stabilized at tree-level via R-R and
$H_3$ form fluxes. 
In \cite{Lust:2019zwm}, it was noted that the DGKT models
satisfy the ADC, but not in its strong version. 
Let us understand this claim through a particularly simple yet enlightening example.
The compactification manifold we choose is the isotropic six-torus $T^6$, 
which means that the moduli superfields will be identified as 
$T_1=T_2=T_3 \equiv T$ and $U_1=U_2=U_3\equiv U$. So we have only
three remaining superfields $S, T, U$. 
The real parts of the superfields are defined as
\eq{
          \tau=r_x r_y\,,\qquad   s=e^{-\phi} r_x^3\,,\qquad
          u=e^{-\phi} r_x r_y^2\,.
}
The K\"ahler potential is as usual given by
\eq{
       K=-3 \log(T+\ov T)-3 \log(U+\ov U)-\log(S+\ov S)\,,
}
while for the superpotential we make the restricted choice
\eq{
           W = if_0 T^3 - 3if_4 T + ih_0 S + 3ih_1 U\,,
}
stabilizing all axions at vanishing value. The only non-vanishing
contributions to the $C_7$-form tadpole come from the combinations
$f_0 h_i$ that in a full model need to be cancelled by the
$O6$-planes and $D6$-branes. Thus, the four-form flux $f_4$ is not
bounded by any topological condition.

Computing the scalar potential, one realizes that
there exist both supersymmetric and non-supersymmetric AdS minima. 
For instance in the supersymmetric vacuum, the moduli are stabilized at
\eq{
                 \tau=\kappa {f_4^{1\over 2}\over f_0^{1\over 2}}\,,\qquad
                  s={2\kappa\over 3} {f_4^{3\over 2}\over f_0^{1\over 2} h_0}\,,\qquad
                  u={2\kappa} {f_4^{3\over 2}\over f_0^{1\over 2} h_1}\,
}
with $\kappa=\sqrt{5/3}$. For the non-supersymmetric minima only the
numerical prefactors change. 
The  effective masses of the moduli all scale in same way as
\eq{
                  m_{\rm mod}^2\sim -\Lambda\sim {f_0^{5\over 2} h_0 h_1^3\over
                    f_4^{9\over 2}} M_{\rm pl}^2\,.
}
Therefore the AM-SSC is satisfied, also in its strong form.
Since we have two geometric radii, there will also exist two KK scales. Those are
\eq{
                m_{{\rm KK},1}^2 &= {M_s^2\over r_x^2}={M_{\rm pl}^2\over s\,\tau\,u}={f_0^{3\over 2} h_0 h_1\over
                    f_4^{7\over 2}} M_{\rm pl}^2 \,, \qquad\quad
               m_{{\rm KK},2}^2 = {M_s^2\over r_y^2}={M_{\rm pl}^2\over \tau\,u^2}={f_0^{3\over 2}  h_1^2\over
                    f_4^{7\over 2}} M_{\rm pl}^2     \,.        
} 
Choosing the fluxes as  $f_0, h_0,h_1=O(1)$, $f_4 \gg 1$, the values
of the moduli are in the perturbative regime where the string coupling
is weak ($g_s\ll 1$) and the radii large.
One notices that in this regime the KK scales are parametrically larger than the moduli masses  and one
has
\eq{         
          m_{{\rm KK},i}^2\sim |\Lambda|^{7\over 9}\,,
}
thus satisfying the ADC with $\alpha=7/18$, for both the SUSY and non-SUSY vacua. 
Since $\alpha\neq \frac{1}{2}$, as noted in \cite{Lust:2019zwm} 
the strong ADC is indeed violated.
This mass scaling above reflects  the original claim \cite{DeWolfe:2005uu}
that type IIA  flux compactifications admit a dilute flux limit (where
$f_4\to\infty$).
 
An important question is whether this effective 4D analysis can really
be uplifted to a full solution of the 10D string equations of motion.
The difficulty lies in the proper description of the localized
O6-plane/D6-brane sources.
A couple of important steps in this direction have been performed recently.                 
In \cite{Junghans:2020acz}, the backreaction was taken into account
in a perturbation around the known smeared solution.
Another interesting idea was put forward in \cite{Buratti:2020kda},
where a refinement  of the ADC was proposed that includes 
a discrete $\mathbb Z_k$  symmetry. 
Finally, in \cite{Marchesano:2020qvg} approximate solutions to the 10D supersymmetry equations with similar features to the 4D DGKT vacua were found, satisfying a non-trivial test for the 4D vacua to admit a 10D uplift. All these approaches do not hit any obstacle for the scale separation of DGKT vacua to hold.

\subsection{Example II: Type IIA with geometric flux}

Let us now consider an example which is more similar  to the
Freund-Rubin case in the sense that 
there is also a  geometric (curvature) contribution necessary 
for the stabilization of the moduli. Describing 
the $AdS_5\times S^5$ case from an effective 5D perspective\footnote{
Similar backgrounds of the form 
$AdS_2\times M^{(1)}\times\ldots\times M^{(n)}$, with $M^{(i)}$ Riemannian spaces,
have been examined recently in \cite{Lust:2020npd}.
}, 
a positive 5-form flux contribution is balanced against 
a  negative curvature contribution.
Such backgrounds can be  described by turning on geometric flux
$\omega$ in the effective theory. While many models of this type have 
been considered in \cite{Camara:2005dc}, here we just focus on an
illustrative example.

Like in the previous example, consider now a toroidal type IIA orientifold with the superpotential 
\eq{
           W=f_6 + 3f_2 T^2 - \omega_0 S T -3 \omega_1  U T\,,
}
where the last two contributions arise from geometric fluxes.
The scalar potential  is found to have a supersymmetric AdS minimum at
\eq{
                 \tau={1\over 3} {f_6^{1\over 2}\over f_2^{1\over 2}}\,,\qquad
                  s=2{f_2^{1\over 2}  f_6^{1\over 2}\over  \omega_0}\,,\qquad
                  u=2{f_2^{1\over 2}  f_6^{1\over 2} \over \omega_1}\,,
}
in which the moduli masses scale as
\eq{
                  m_{\rm mod}^2\sim -\Lambda\sim { \omega_0\, \omega_1^3\over
                    f_2^{1\over 2}  f_6^{3\over 2}   } M_{\rm pl}^2\,.
}
In this case the two KK scales are
\eq{
       m_{{\rm KK},1}^2 = {  \omega_0 \,\omega_1\over
                     f_2^{1\over 2}  f_6^{3\over 2} } M_{\rm
                     pl}^2\,,\qquad\quad
       m_{{\rm KK},2}^2 = {  \omega_1^2\over
                     f_2^{1\over 2}  f_6^{3\over 2} } M_{\rm
                     pl}^2
}
which satisfy $m_{{\rm KK},1}^2 \sim \frac{m_{\rm
  mod}^2}{\omega_1^2}$ and $m_{{\rm KK},2}^2 \sim \frac{m_{\rm
  mod}^2}{(\omega_1 \omega_2)}$. Naively looking at this result,
  one would conclude  that a suitable selection of fluxes  allows
  for a parametric scale separation, although in the wrong direction
$m_{{\rm KK}}< m_{\rm mod}$.

However, as anticipated in \cite{Blumenhagen:2019vgj} and concretely  shown in
\cite{Font:2019uva},
 to draw  such a conclusion it is important to take
the backreaction of the fluxes onto the metric into account.
Looking more carefully into the situation, 
one realizes that the actual eigenvalue problem one has to solve is
\eq{
\label{KKback}
                 \Delta_{\ov G} \chi(y)  + \Phi(\ov G,\ov F,\ldots)\,   \chi(y)  =-m^2_{\rm KK}\, \chi(y)
}
where $\ov G,\ov F$ denote the fully backreacted background
metric and flux. 
When doing so, the geometric fluxes in the denominator also cancel and parametrically one
indeed finds $m_{{\rm KK}}^2\sim  m_{\rm mod}^2 \sim |\Lambda|$.
Therefore, in such models with geometric flux there is no
parametric separation of the KK scale and the moduli masses.
Note that this means that both in the supersymmetric and
non-supersymmetric case the relation from the strong ADS conjecture holds.

Thus, a well motivated guess is that  the backreaction is of paramount
importance whenever geometric effects, such as geometric fluxes,
contribute to the moduli stabilization. Moreover, we would like to
suggest
that the relevant criterion for the strong version of the  ADS
conjecture ($\alpha=1/2$) to hold  is not 
supersymmetry  but whether  geometric
fluxes are involved when describing the model  from the 4D effective field theory perspective.

\section{Issues of strongly warped KKLT}
\label{sec_KKLT}

The KKLT scenario \cite{Kachru:2003aw} has been proposed to be a controlled 
set-up of string moduli stabilization that eventually leads
to a dS minimum. It is  not a fully fledged string theory construction
but rather consists of a number of constituents  that were argued to 
be generically present in consistent string models,  working together
in an intricate manner to give dS vacua.

Here we do not review the full KKLT scenario, but rather remind the
reader of the essential steps. For more details we refer to the literature.
One is working in an effective 4D supergravity theory that results 
from a compactification of the type IIB superstring with fluxes,
branes and orientifold planes. Then one proceeds in three steps:

\begin{enumerate}
\item{Stabilize  the complex structure and axio-dilaton moduli via
three-form fluxes in a no-scale non-supersymmetric Minkowski minimum with $W_0\ll 1$.}
\item{Consider the effective theory of the light K\"ahler
    modulus  described by a K\"ahler potential and superpotential
\eq{
                 K=-3\log(T+\ov T)\,, \qquad   W=W_0 + A e^{-aT}\,.
} 
The non-perturbative effect conspires with the tiny value of $W_0$ to
stabilize $T$ in a supersymmetric AdS minimum.}
\item{Uplift the minimum to dS via $\overline{D3}$  branes at the
    tip of a warped throat.}
\end{enumerate}

\vspace{0.0cm}
This set-up has been scrutinized from various sides in the past. First
the validity of step 3  was questioned, namely whether an 
$\overline{D3}$-brane at the tip of a
warped throat is really a stable configuration 
(see \cite{Danielsson:2018ztv} for a review,
and \cite{Bena:2018fqc,Bena:2019sxm,Dudas:2019pls} for recent developments). 
Moreover, it has been
questioned whether the  4D description of the KKLT AdS minimum does
really uplift to a full   10D  solution of string theory
\cite{Moritz:2017xto,Kallosh:2018wme,
  Kallosh:2018psh,Gautason:2018gln,Hamada:2018qef,Hamada:2019ack,
  Carta:2019rhx,Gautason:2019jwq,Bena:2019mte}.
More recently, there were also growing concerns even about step 1.

One needs $W_0\ll 1$ for the KKLT construction\footnote{
It was recently argued that large $W_0$ values can also provide the
supergravity potential with a dS minimum \cite{Linde:2020mdk}. It is however
not clear that this supergravity solution uplifts to a solution of string theory.
}, so this should better  not be in the
swampland. Recently, in the large complex structure regime, a mechanism has
been described that gives $W_0=0$ at leading order where
subleading (instanton like) terms provide the stabilization of a
 final light complex structure modulus \cite{Demirtas:2019sip}. This leads
to exponentially small values of $W_0$. Clearly, for the later uplift
one actually needs a similar controllable mechanism \cite{ABBS} close to a
conifold point in the complex structure moduli space  where large
warping can occur.

Relatedly, for the uplift one has to invoke an effective action that is
valid in the strongly warped regime. Based on earlier work
\cite{Douglas:2007tu}, 
this question has been addressed  recently 
\cite{Bena:2018fqc,Bena:2019sxm,Dudas:2019pls,Randall:2019ent}
and in the following we will review some of the main results, in particular
of \cite{Blumenhagen:2019qcg}. For more details we refer the reader 
to the original paper\footnote{For related recent work on analyzing warped
  throats and their relation to swampland conjectures see \cite{Blumenhagen:2016bfp,Hebecker:2018yxs,Buratti:2018xjt}.}.

\subsection{Effective action in the strongly warped regime}

Strongly warped throats can develop close to a conifold point $Z=0$ in the complex structure moduli space. 
The ultimate question is whether 
the effective action for the conifold modulus $Z$ is really under
control.

There exists a proposal for the effective action of the conifold
modulus $Z$ and the warped volume modulus ${\cal V}_w$ in the
strongly warped regime \cite{Douglas:2007tu}. This is supposed to be valid
at a point very close to a conifold singularity where warping effects
become relevant.
At such points the geometry develops a long throat,
at the tip of which a three-cycle $A$ becomes very small. 
The corresponding complex structure modulus is defined as $Z=\int_A
\Omega_3$, where the physical size of the 3-cycle $A$ is ${\rm
  Vol}(A)={\cal V}^{1\over 2} |Z|$. Therefore, in the regime ${\cal V}
|Z|^2\gg 1$ warping can be neglected whereas for ${\cal V}
|Z|^2\ll 1$ it is substantial. Moreover, such a warped throat can be supported
by turning on three-form fluxes $M=\frac{1}{(2\pi)^2\alpha'}\int_A F_3$ and $K=\frac{1}{(2\pi)^2\alpha'}\int_B H_3$.

\begin{figure}[ht]
	\vspace{-2ex}
	\begin{tikzpicture}
	\tikzset{
		shadowY/.style={preaction={transform canvas={shift={(-0.3pt,-0.1pt)}},draw=gray,semithick}},
		shadowS2/.style={preaction={transform canvas={shift={(-0.1pt,-.1pt)}},draw=gray,semithick}},
		shadowS3/.style={preaction={transform canvas={shift={(0pt,-0.3pt)}},draw=gray,semithick}},
	}
	\node[anchor=south west,inner sep=0] (modulispace) at (0,0) {\includegraphics[width=0.8\textwidth]{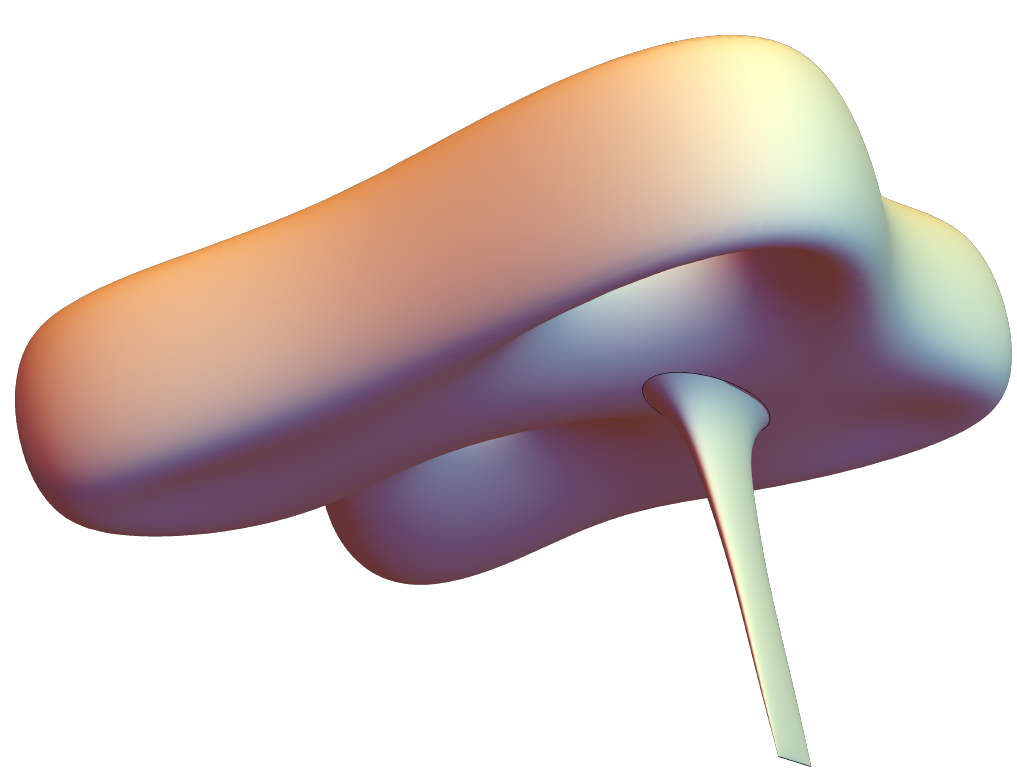}};
	\draw[->,shadowY] (9.2,0.2) -- ($ (9.15,0.2) + (104:1) $);
	\draw[->,shadowS2] (9.2,0.2) -- ($ (9.15,0.2) + (215:0.6) $);
	\draw[->,shadowS3] (9.2,0.2) -- ($ (9.15,0.2) + (-17:0.9) $);
	\draw[-,semithick] (2,6.7) -- (5,6);
	\draw[-,semithick] (7,2) -- (8.4,3.2);
	\node(y) at ($ (9.15,0.2) + (107:1.2) $) {\footnotesize $y$};
	\node(S2) at ($ (9.15,0.2) + (210:0.84) $) {\footnotesize $S^2$};
	\node(S3) at ($ (9.15,0.2) + (-14:1.15) $) {\footnotesize $S^3$}; 
	\node(bulk) at (1.4,6.75) {bulk};  
	\node(throat) at (6.5,1.8) {throat};
	\draw [decorate,decoration={brace,amplitude=8pt},semithick,xshift=-4pt,yshift=0pt]
	($ (9.30,4.4) + (10:0.03) $) -- ($ (9.9,0.17) + (10:0.03) $) node [black,midway,xshift=21pt] {\footnotesize
		$y_{UV}$};
	\end{tikzpicture}
	\caption{A sketch of a Calabi-Yau with a KS-throat. At the tip of the throat the $S^2$ shrinks to zero size while the $S^3$ remains finite. $y_{UV}$	marks the cutoff where the throat meets the bulk.} 
	\label{fig_throat}
\end{figure}
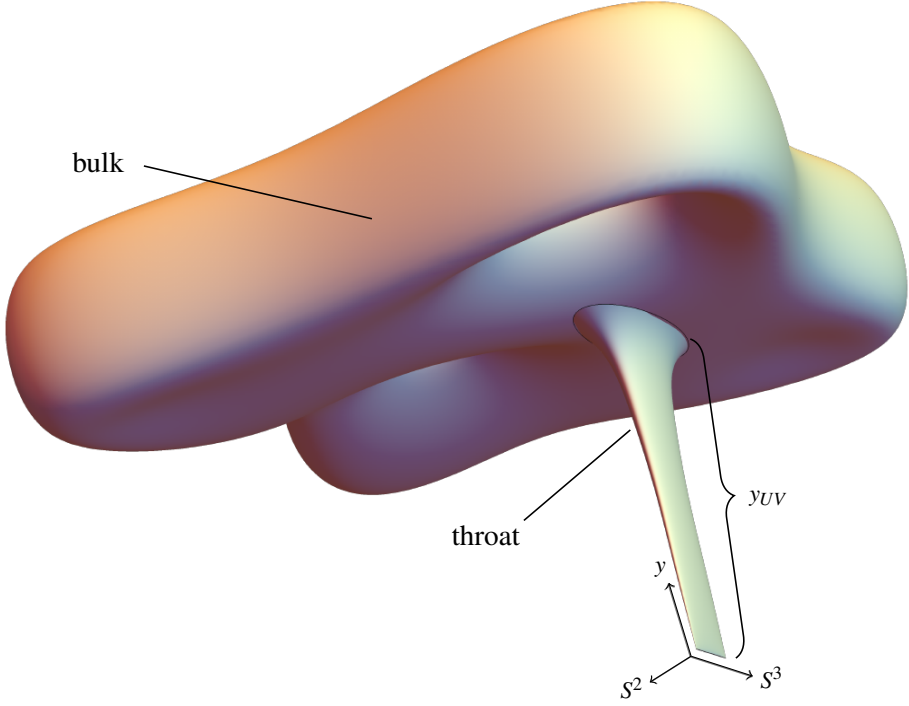

Locally this deformed conifold geometry is described by a
Klebanov-Strassler (KS)
throat  \cite{Klebanov:2000hb}, for which the metric is explicitly known. One can think
of the total geometry as such a KS throat of  length $y_{\rm UV}$
glued to  the bulk Calabi-Yau manifold. The warp factor along the
throat direction $y$ is given as
\eq{
\label{warpfactor2}
              e^{-4A({y})}\approx  {g_s M^2\over
                ({\cal V}_w |Z|^2)^{2\over 3}} {\cal
           I}(y) \,
}
where ${\cal V}_w |Z|^2\ll 1$ as mentioned above and ${\cal I}(y)$ 
is an explicitly known function in the KS solution.
We have depicted a sketch of the warped geometry in figure \ref{fig_throat}.

\vspace{0.3cm}

Now,  it is assumed that all other complex structure moduli measuring the size of
the bulk geometry are stabilized by fluxes at a larger mass scale so that the
lightest moduli are $Z$ and the overall K\"ahler modulus $T$.
It turns out that the  relevant parameters are $\big\{ g_s, M, y_{\rm
  UV} \big\}$. To be in the large radius regime they need to satisfy
$1\ll g_s M^2$ and  $1\ll g_s M^2\,  y_{\rm UV}^2$ .
It was shown in \cite{Douglas:2007tu} that the K\"ahler potential for $Z$ and $T$ reads
\eq{ 
\label{kaehlerpotb}
    K=-3\log(T+\ov T) +  {c'  g_s M^2 \vert Z\vert^{2\over
      3}\over (T+\ov T)} + \ldots
} 
featuring a  no-scale structure (up to $O(g_s M^2)$).
The three-form flux induced superpotential for the conifold modulus
$Z$ reads  
 \eq{
\label{superpotb}
     W=w_{0}-{M\over 2\pi i} Z \log Z  +i  {K S} Z  +O(Z)\,,
}
where as above  $M,K$ denote the fluxes through the $A,B$ cycles and
$w_0$ denotes a constant contribution to the superpotential. This will 
in general be a non-zero complex number and as mentioned, it would be interesting to generalize the
mechanism for $W_0\ll 1$  of \cite{Demirtas:2019sip} to the conifold regime. 
Self-consistently, the above supergravity model fixes the conifold modulus at
the exponentially small value
\eq{
|Z_0|\sim \exp\left(-{2\pi K\over g_s M}\right)\,
}
and its mass turns out to be 
\eq{
\label{massconi}
                m^2_Z\simeq  {1\over  g_s M^2} \left(
                  |Z|\over  {\cal V}_w \right)^{2\over 3}  M_{\rm pl}^2
\simeq   {\big({\cal
                      V}_w \vert Z\vert^2\big)^{1\over 3}\over g_s^{3/2}\, M^2}
                  M_{\rm s}^2 \,.
}
This nicely shows that in the regime of strong warping   $ {\cal
  V}_w \vert Z\vert^2\ll 1$ one indeed finds  $m_Z<M_s$.
Moreover, we note that this mass is exponentially small so the
question arises whether it is really heavier than the K\"ahler modulus that is
stabilized in step 2.

Computing the scalar potential and adding the contributions of 
an  $\ov{D3}$ brane  at the tip of the throat one obtains
\eq{
\label{Vantitot}
V_{\rm tot}= {9\over 2 c' M^2} {|Z|^{4\over 3}\over {\rm Re}(T)^2} \Big[
      \Big({\textstyle {M\over 2\pi}}\log |Z|+{\textstyle {K\over g_s}}\Big)^2 +
      {c' c''\over g_s} \Big]\,.
}
One observes that  the three-form flux and the $\ov{D3}$ brane contribution
scale in the same way with the moduli $Z$ and $T$.
Therefore,  it is not justified to first integrate out $Z$ and then
perform the uplift. Plugging in concrete numbers for the constants
$c',c''$, it was shown \cite{Bena:2018fqc}
that the minimum continues to exist only for $\sqrt{g_s} M > 6.8$.
Thus, in the perturbative regime ($g_s<1$) one needs a large flux
which could be  in conflict with tadpole cancellation.

Now, let us consider step 2 of KKLT and add the non-perturbative term $W_T=A\exp(-aT)$.
Let us assume that one can fix the other complex structure moduli and the
axio-dilaton such that $w_0$ and $Z$ are of the same order. More
on this will be reported in \cite{ABBS}. Then the  scale  of the superpotential in the
minimum is 
\eq{
|W_0|\sim |Z_0|\sim \exp\left( -{2\pi  K\over g_s M}\right)
}
and therefore also exponentially small. As a consequence, in the KKLT AdS minimum, the mass of $T$ scales as
\eq{
          m^2_T\sim {W_0^2\over {\cal V}_w^{2/3}} M_{\rm pl}^2 <
          m^2_Z \,.
}
Therefore, self-consistently the mass of the K\"ahler modulus is still
lighter than the mass of the conifold modulus.
It seems that everything is fine once $|w_0|\sim |Z|$ can indeed be
realized in the string landscape.

\subsection{KK modes and emergence}

However, there is one issue that requires closer inspection, namely
that  in the strongly warped throat
geometry one direction (namely $y$ along the throat) becomes
much larger than the other bulk directions. Therefore, one is
dealing with a highly non-isotropic geometry that could support
very light Kaluza-Klein modes \cite{Frey:2006wv, Burgess:2006mn,Shiu:2008ry,deAlwis:2016cty}.
We will see that such modes indeed
exist and that, reminiscent  of infinite distance limits their mass scale is
related to the distance of the conifold point in the complex structure
moduli space via emergence \footnote{In a similar spirit,  in \cite{Enriquez-Rojo:2020pqm} the WGC, the
  swampland distance conjecture and emergence have been analyzed
for type IIB orientifolds with closed string abelian gauge fields.}.

To estimate the  mass scale of the KK modes one needs to 
solve the  six-dimensional warped Laplace equation
\eq{
\label{laplacewarp}
        e^{4A(y)} \nabla^m \nabla_m \chi(y) - m^2\, e^{2A(y)}
        \chi(y)=-m^2_{\rm KK}\, \chi(y)\,.
}
In the regime close to the tip of
the throat  one can use the explicitly known KS solution and 
it has been shown in \cite{Blumenhagen:2019qcg} that  there exist light KK modes  with mass
\eq{
                   {m^2_{\rm KK}} = c \, {n^2\over
                     y_{\rm UV}^2}\, m_Z^2 \,.
} 
With respect to $g_s$,  $M$ and $({\cal V}_w |Z|^2)$,
this scales precisely as $m^2_Z$. The requirement that the description
in terms of a warped throat is consistent constrains the UV cutoff
$y_{\rm UV}$ to $y_{\rm UV}>1$. In this case there are a finite number
of KK modes which are lighter than the conifold modulus. Thus the
description in terms of the effective action could break down. 

Remarkably, the existence and mass scale of these KK modes is consistent with the picture of
emergence of the field space metric \cite{Grimm:2018ohb}, even
though here the conifold locus is at finite distance in the complex
structure moduli space.
Indeed, the distance from the conifold locus in the warped case in terms of the $Z$ modulus is given by
\eq{
            \Phi=\int d\zeta \sqrt{G_{Z\ov Z}}\sim
            \left( {|Z| \over {\cal V}_w}\right)^{\! 1\over 3} \;. 
}
Guided by the emergence proposal, one can determine the cut-off
$\Lambda$ of the effective theory.
For that purpose, one  assumes that in the regime $g_s M^2\gg 1$ the tower of KK modes 
\eq{
   \Delta m \approx {1\over \sqrt{g_s M^2} \, y_{\rm UV}} \left({|Z|\over {\cal
          V}_w}\right)^{\!{1\over 3}} M_{\rm pl}
}
is lighter than the cut-off. When these are integrated out, the field
space metric should get a one-loop correction proportional to the tree-level
result. Assuming ${\rm deg}(n)=1$ one finds 
\begin{equation}
\label{eq:KK1loop}
	\begin{aligned}
		 g_{Z\ov Z}^{\rm 1-loop}
                  &\sim N_{\rm sp}^3\frac{1}{g_s M^2y_{\rm UV}^2}\frac{1}{(\mathcal{V}_w|Z|^2)^{2/3}}\,.
	\end{aligned}
\end{equation}
Matching this  with  the tree level result
\begin{equation}
g_{Z\ov Z}^{\rm tree}\sim \frac{g_sM^2}{(\mathcal{V}_w |Z|^2)^{2/3}}\,,
\end{equation}
one can fix the species scale $N_{\rm sp}$, i.e. the number of KK-states below the cutoff, as
\begin{equation}
	N_{\rm sp}\sim\left(g_s M^2 y_{\rm UV}\right)^{2/3}\,.
\end{equation}
Then, as one knows the spacing and number of modes below the cutoff, one finds for the  cut-off
\eq{
    \Lambda\sim \sqrt{g_s M^2} \left({|Z|\over {\cal
          V}_w}\right)^{\!{1\over 3}} M_{\rm pl}\,.
}
Intriguingly, this cutoff equals the mass of a D3-brane wrapping the A-cycle $S^3$
which vanishes at the conifold. 
Indeed, it is well known that the conifold singularity in the complex
structure moduli space arises from integrating out precisely such a
D3-brane \cite{Strominger:1995cz}. Including the warp factor,
one can determine the mass of the wrapped D3-brane as
\eq{
	\label{masswrappedd3}
	m^2_{\rm D3}\sim g_s^{1\over 2} M^2  ({\cal V}_w
	|Z|^2)^{1\over  3} M_s^2\sim
	g_s M^2  \left( {|Z|\over {\cal V}_w}\right)^{\!{2\over  3}}
        M_{\rm pl}^2\sim \Lambda^2\,.
}
How does this cut-off relate to the
earlier cut-off $y_{\rm UV}$ of the throat length? An upper bound for
 $y_{\rm UV}$ can be obtained by demanding the throat volume ${\cal
  V}^{\rm throat}_w$ to be smaller than the volume of the whole Calabi Yau, ${\cal V}^{\rm throat}_w<{\cal V}_w$.
At leading order in the Klebanov-Strassler solution this constraint leads to
\eq{
	y_{\rm UV} \lesssim 3\log\left( {M_{\rm pl}/ \Lambda}\right)\,,
}
relating the two introduced cut-offs $y_{\rm UV}$ and $\Lambda$ in a
simple manner.

In summary, the following picture for the mass scales in the effective theory of the warped throat arises:
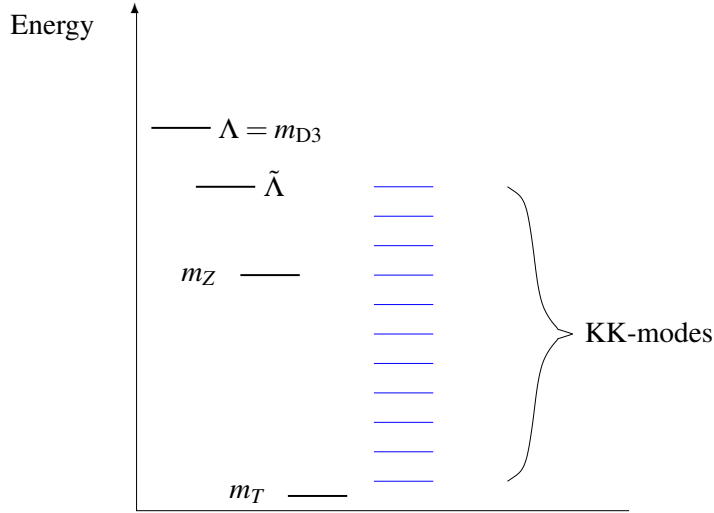
\begin{figure}[ht]
  \centering
    \includegraphics[width=6.5151405cm]{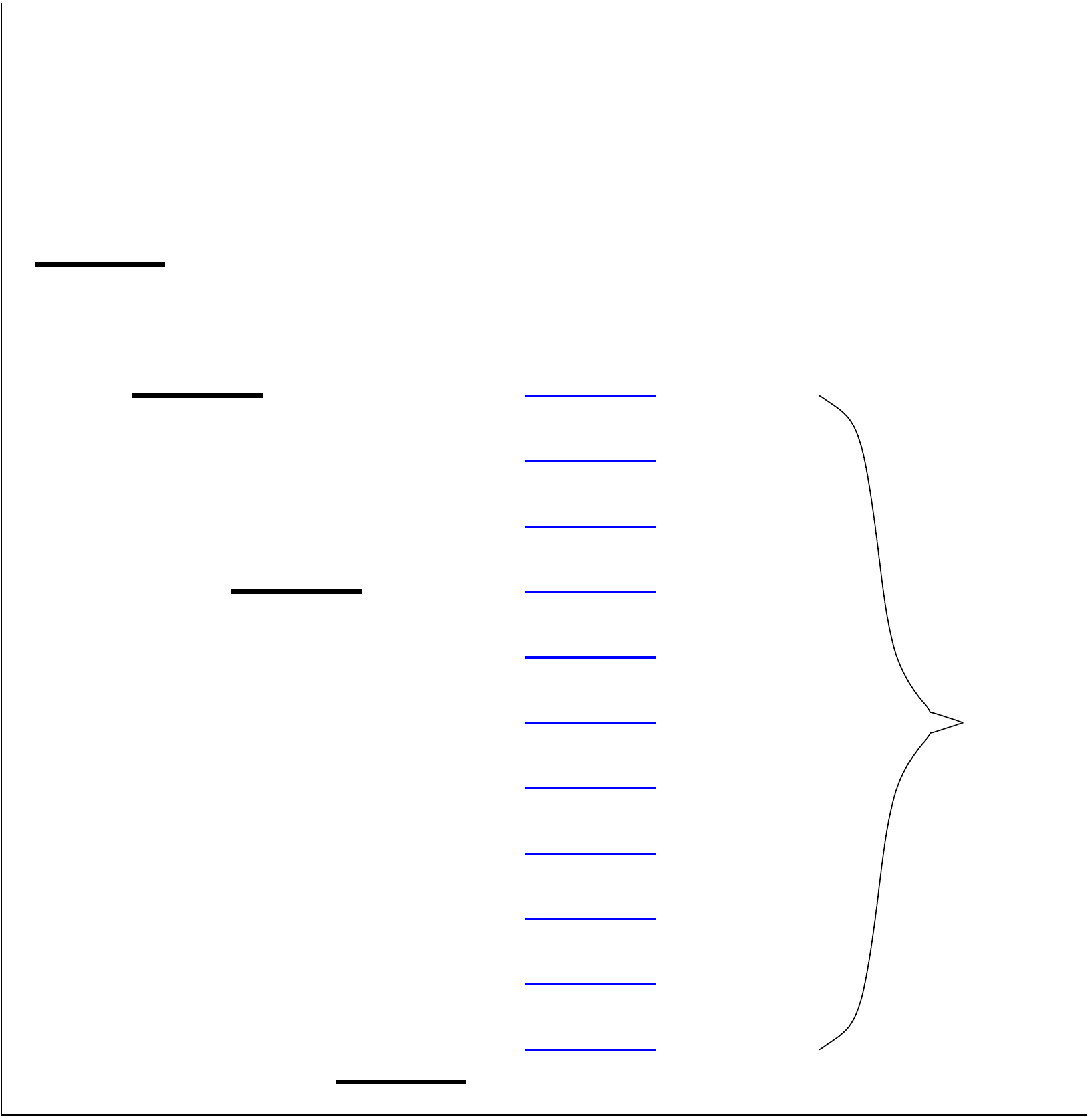}
\begin{picture}(0,0)
    \put(-157,141){$\Lambda=m_{\rm D3}$}
        \put(-140,119){$\tilde\Lambda$}
        \put(-171,87){$m_{Z}$}
        \put(-153,5){$m_{T}$}
        \put(-20,64){KK-modes}
        \put(-235,180){Energy}
        \put(-188.5,192){\vector(0,1){0}}
    \end{picture}
  \caption{Mass scales in the warped throat in the regime $y_{\rm UV}>1$.}
  \label{fig:spec}
\end{figure}

\noindent
As in the swampland distance conjecture, there are KK-modes which are
lighter than the cut-off so that one expects
the effective theory to be  outside of the controlled regime. 
But the emergence proposal implies that the corrections induced by
these modes  are proportional to the tree level
result, only changing numerical factors. 
It is not clear to us what this extra special structure implies
for the reliability of the ``naive'' effective action.
Could this not indicate that it is  possible that 
the  ``naive'' effective action is much better than one might
initially think?
For sure the effect of including the extra light KK modes in the action
will not change the functional dependence of the kinetic term of the modulus $Z$, even though there
might be other more drastic effects. Unfortunately,  we cannot give a conclusive
answer to the question of the reliability of the effective action
and leave it for  future work.

\section{Quantum Log-Corrections to Swampland Conjectures}
\label{sec_quantumLog}

Having reviewed recent work on  the effective theory in the warped throat,
let us now focus on the KKLT AdS minimum, while assuming that $W_0$ can indeed be 
tuned to exponentially small values. Since the masses of the moduli $T,Z$ parametrically
satisfy $m_T<m_Z$, we are confident that the AdS minimum where $T$ is stabilized 
by a  non-perturbative contribution, indeed belongs to the string theory landscape. 
In the following we will review the arguments
\cite{Blumenhagen:2019vgj} suggesting  that  $\log$-corrections to 
the AdS/moduli scale separation and the AdS distance
conjecture occur.

We have already seen in section \ref{sec_one} that the quantum
generalization  of the dS swampland
conjecture, namely the TCC, also included such
$\log$-corrections. In this case,  the bound on the life-time 
of a dS phase got weakened \eqref{TCClifetime}.
Therefore, it is natural to wonder whether a similar relaxation of the
bounds also occurs  for the AdS conjectures.

\subsection{KKLT and the  AdS-moduli scale separation conjecture}

Recalling once more the KKLT construction, the AdS vacuum occurs in its second step. 
Assuming real values for $W_0,\,A\in\mathbb{R}$ with $W_0 \,A<0$, the scalar
potential after freezing the axion at $\theta=0$ reads
\eq{
\label{hannover96}
                  V_{\rm KKLT}={a A^2 \over 6\tau^2} e^{-2
                    a\tau}(3+a\tau) +{a A 
                    W_0\over 2\tau^2} e^{-a\tau}\,.
}
In order to determine the position of the supersymmetric AdS vacuum, 
one needs to solve the transcendental equation
\eq{
\label{KKLTmineq}
               A(2a\tau+3)=- 3W_0 e^{a\tau} \,,
}
leading to a negative cosmological constant 
\eq{
\label{KKLTLam}
                 \Lambda= -{a^2 A^2 \over 6\tau} e^{-2a\tau} \,.
}
Concerning the AdS distance conjecture, it is clear that the $\Lambda\to 0$ limit occurs for $\tau\to\infty$,
i.e. at infinite distance in the moduli space.
One can easily compute the mass of the $\tau$ modulus in the minimum $V_0$ as
\eq{
\label{KKLTmodmass}
	m_{\tau}^2 = \frac{1}{2} K^{\tau\tau} {\partial_\tau}^2 \,V \big|_{V_0}
		={a^2 A^2  \over 9\tau}(2+5 a\tau+2a^2 \tau^2)\,e^{-2a\tau}\, .
}           
This is the modulus relevant for the AdS-moduli scale separation conjecture. Even taking into account the conifold modulus and the additional KK modes of the throat, the K{\"a}hler modulus remains the lightest.
Plugging in equation \eqref{KKLTLam}, we can express it as
\eq{
                     m^2_{\tau}=  -{2\over 3} (2+5
                     a\tau+2a^2 \tau^2)\, \Lambda \,.
}
It is clear that the $\tau$-dependent terms inside the parenthesis
violate the bound  \eqref{mscalesep}, so  in the controlled regime
$a\tau\gg 1$ scale separation is possible to some extent.
Expressing the above equation only in terms of $\Lambda$ and 
neglecting $\log\log$-corrections,   one can write
\eq{
\label{KKLTscaling}
      m^2_{\tau}=  - \Big(c_2^2 \log^2(-\Lambda)+c_1 \log(-\Lambda) +  c_0\Big)
      \, \Lambda \,.
}
This suggests to declare a quantum version of the  AdS/moduli scale
separation conjecture taking the form
\eq{
                    m_{\rm mod} \,R_{\rm AdS} \le c \log( R_{\rm AdS} M_{\rm pl})\,
}
where we have reintroduced the Planck-scale.
This gives a weaker bound on the combination on the left-hand side
allowing a $\log$-type scale separation between the lightest moduli
mass and the AdS scale. Moreover, in the $M_{\rm pl}\to \infty$ limit where gravity decouples,
the quantum version of the conjecture becomes trivial, as expected.


\subsection{KKLT and the  AdS distance conjecture}

Let us  now discuss the AdS distance conjecture, according to which in
the $\Lambda\to 0$ limit  there must appear a tower of states with
masses  $m_{\rm tower}=c |\Lambda|^\alpha$.
In the original paper \cite{Lust:2019zwm} it was not clear whether the
KKLT AdS vacuum abides by this conjecture.
The reason is that it is not trivial to identify which is the tower that should behave in the above way.
The tower that would immediately come to mind is the Kaluza-Klein tower.
Usually, the mass scale of the KK states on an isotropic manifold is estimated
as the inverse of the relevant length scale. For the KKLT case, that would give the estimate
\eq{
\label{KKnaiveKKLT}
                    m_{\rm KK}\sim {1\over \tau}\,.
}
However, this does not exhibit the behaviour expected by the ADC, 
regardless of the value of the exponent in the conjecture. In fact, 
this mass scale is exponentially larger than $|\Lambda|^{\alpha}$ for any $\alpha$.
This exponential mismatch however should not necessarily  be interpreted as a
failure of the conjecture, but rather as a lack of knowledge of the
relevant tower of states or the reliability of the naive estimate
\eqref{KKnaiveKKLT}.

Concerning the latter possibility, we have already discussed in
section \ref{sec_flux} that the solution to the actual
eigenvalue problem \eqref{KKback} can have  important effects.
Coming to the other point, it is not necessary that the naive  estimate for
the KK scale \eqref{KKnaiveKKLT}  really gives the mass of all
possible  KK modes. This is even evident from the example of a
non-isotropic two-dimensional torus with a short and a long circle.
The long circle gives rise to a KK tower whose mass is lower than \eqref{KKnaiveKKLT}. 
The strongly warped throat geometry is precisely of this type and there
we have identified already  a tower of lighter KK modes localized close to the tip of the throat, scaling as
\eq{\label{warpedKKmass}
                m^2_{\rm KK}\sim {1\over y_{\rm UV}^2} \left( {Z\over
                    {\cal V}_w}\right)^{2\over 3}\,. 
}
In the limit where  the throat just fits into the warped CY we can approximate
\eq{\label{yUV}
                   y_{\rm UV}\sim -\log\left( {|Z|/ \tau^{3\over
                         2}}\right)\sim \tau\,,
}
so that there  actually exists  an exponentially light KK tower.
Its mass scale can be expressed at the KKLT AdS minimum as
\eq{
\label{KKtrueKKLTb}
                  m^2_{\rm KK}\sim {1\over \tau^2}\,  {e^{-{2\over 3}a\tau}\over
                \tau^{1\over 3}}\sim {1\over {\log^2( -\Lambda)}} |\Lambda|^{1\over 3}\,.
}
One realizes that this behavior, up to the $\log$-corrections,
satisfies the ADC for  $\alpha=1/6$.  Since $\alpha<1/2$, one has scale
separation between the AdS scale and the size of the internal
geometry.

\subsection{KKLT and emergence}

Let us now comment on an intriguing  connection to the emergence proposal.   
Recall from section \ref{sec_KKLT} that  the subleading term in the
K\"ahler potential
\eq{
		K=-3\log(T+\ov T) +  c  {|Z|^{2\over 3}\over (T+\ov T)}\,
}
is corrected by the tower of red-shifted KK modes with mass scale \eqref{warpedKKmass} 
at the tip of the throat. For the species scale one finds
\eq{
		\Lambda_{\rm sp}^3\sim {|Z|\over \tau^{3\over 2} y_{\rm
		UV}} M_{\rm pl}^3\,,               
} 
where $y_{\rm UV}$ is given in large throat limit by \eqref{yUV}.
Since the  first term in $K$ is also present in the unwarped case, from the 
perspective of the emergence proposal one expects it to 
emerge from integrating out the tower of  bulk KK modes
$\Delta m_{\rm KK,h}\sim 1/\tau$.
Stabilizing $\tau$  via KKLT  and as usual  assuming 
\eq{
		|Z|\sim |W_0|\sim \tau e^{-a\tau}\,,
}
the species scale can be expressed as 
\eq{
		\Lambda_{\rm sp}^3\sim  {e^{-a\tau} \over \tau^{3\over 2}}
		M_{\rm pl}^3\sim  \Lambda^3_{\rm SQCD}\,,
}
where $\Lambda_{\rm SQCD}$ is the dynamically generated mass scale of the confining 
SYM theory which undergoes gaugino condensation,
\eq{
   		{\Lambda^3_{\rm SQCD}= e^{-a/g^2}   M^3}\,.
}
Here  the UV cut-off $M$ of the gauge theory is in turn not 
the Planck scale but rather $M\sim g M_{\rm pl}$ as implied by the 
magnetic weak gravity conjecture \cite{ArkaniHamed:2006dz}.
We think that this match with the expected cut-off of the KKLT model
 provides a compelling result as it relies on a non-trivial combination
of results from section \ref{sec_KKLT} with swampland conjectures.

\subsection{KKLT and the  dS swampland conjectures}

It was already pointed out in \cite{Bedroya:2019snp} that the uplifted KKLT dS minimum does
violate not only the dS swampland conjecture but also
the trans-Planckian censorship conjecture. The reason is simply that the
life-time of the dS vacuum as computed via the Coleman-De Luccia bubble
nucleation \cite{Kachru:2003aw} comes out  exponentially large and violates
\eqref{TCClifetime}.

\begin{figure}[ht]
\begin{center} 
\includegraphics[width=0.37\textwidth]{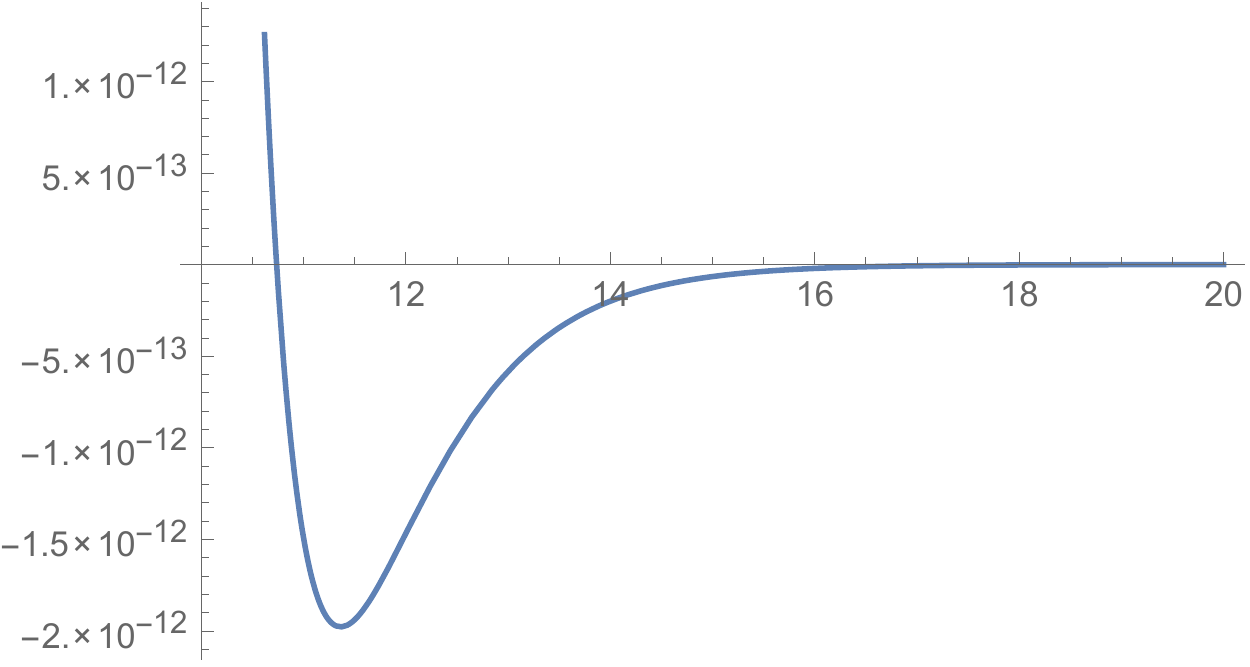}
\begin{picture}(0,0)
  \put(0,50){$\tau$}
   \put(-138,90){$V$}
\end{picture}
\hspace{0.5cm}
\includegraphics[width=0.37\textwidth]{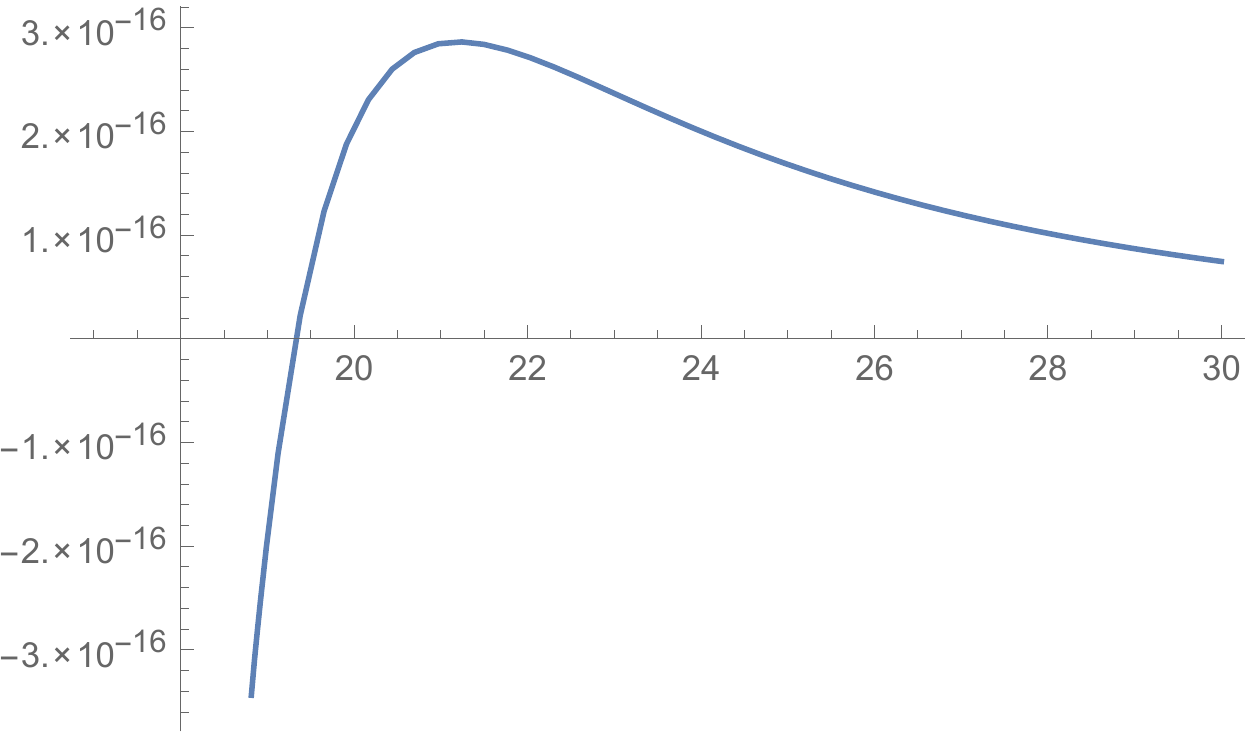}
\begin{picture}(0,0)
  \put(0,50){$\tau$}
   \put(-142,95){$V$}
\end{picture}
\hspace*{15pt}
\end{center} 
\vspace{-5pt}
\caption{The $\alpha'$ corrected KKLT potential (after integrating out
  the axion) for $W_0=10^{-4}$,
  $A=-1$, $a=1$ and $\xi=1$: The figure on the left shows the existence of the AdS
  minimum, while the one on the right reveals the existence of a dS
  saddle point at a larger value of $\tau$.}
\label{fig_ds}
\end{figure}
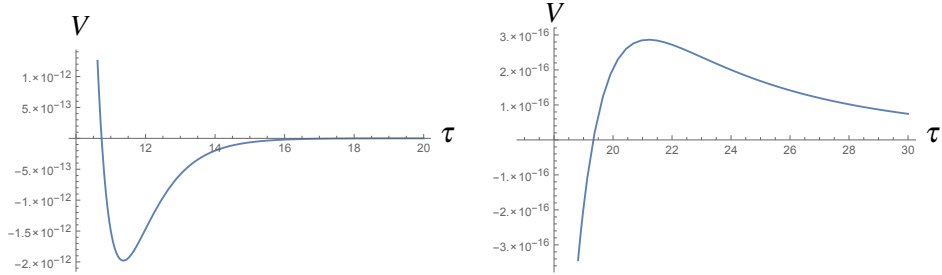

Let us therefore look for a dS saddle point.
Indeed, it is known \cite{Conlon:2018eyr} that including the $\alpha'$ correction to the K\"ahler
potential
\eq{
                K=-2\log\left(  (T+\ov T)^{3\over 2}+{\xi\over
                    2}\right)\,,\qquad
                W=W_0 +A e^{-aT}
}
for $\xi>0$ leads to a dS saddle point at large values of $\tau$.
This is shown in figure \ref{fig_ds}.
\vspace{0.2cm}

\vspace{0.2cm}
\noindent
In the large $a\tau$ regime and after minimization  with
respect to the axion,  the scalar potential for $\tau$ can be approximated
as 
\eq{\label{Vapprox}
             V={3\xi |W_0|^2\over 64 \sqrt{2} \tau^{9\over 2}} - { a |A
             W_0| \over 2 \tau^2} e^{-a\tau} + \ldots
} 
where we neglected subleading terms.
The  position $\tau_0$ of the saddle point  is given by the solution
of  the transcendental relation
\eq{
     |W_0| = {64 \sqrt{2} \,a |A|\over 27 \xi} \tau_0^{5\over 2} (2 + a \tau_0) \,e^{-a\tau_0}\,.
}
Working in the regime $a\tau\gg 1$, the value of the potential at the dS saddle point and its tachyonic
mass can be determined as
\eq{
     V_0\approx {64 \sqrt{2}\over 243 \xi}  a^4 |A|^2\, \tau_0^{5\over 2}\,
     e^{-2a\tau_0}  \,,\qquad
     m_0^2\approx  -{64 \sqrt{2}\over 81 \xi}  a^5 |A|^2\, \tau_0^{7\over 2}\, e^{-2a\tau_0} 
}
so that 
\eq{
                -{m_0^2\over V_0}\approx 3a\tau_0 \approx {3\over 2}\log\left( {1\over
                    V_0} \right)\,.
}
In the regime of control $a\tau\gg 1 $,  this clearly satisfies the second condition in the refined dS
swampland conjecture \eqref{dswamp}. Moreover, by using  $V_0\sim
H^2$, the life-time of the dS saddle point can be expressed
as
\eq{
            T\sim {1\over H \log\left( {M_{\rm pl}\over H} \right)} < {1\over H}<{1\over H} \log\left( {M_{\rm pl}\over H} \right) \,.
}
Therefore, it satisfies the TCC and also contains a $\log
H$-correction. However, in this case the $\log$- factor shortens the
life-time and does not extend it as for the upper bound in the TCC \eqref{TCClifetime}.
This simple example shows that in  non-perturbative KKLT like vacua
indeed $\log$-corrections appear, but they go in the opposite  direction 
than required for a saturation of the TCC bound.

\section{Conclusions}

In this article we have reviewed a couple of issues that arose
when confronting the KKLT scenario with the swampland conjectures.
One not yet excluded possibility could still be that the KKLT scenario
as a whole lies in the swampland of string theory and only
combines string theory ingredients in a manner that can never be
achieved in a  fully fledged string theory construction.

However, we think that so far the KKLT scenario and in particular
the initial AdS minimum has survived  a couple of attacks, that 
even shed some new light onto the many delicate issues hidden in
this simple initial idea. Examples are the questions whether $W_0\ll 1$
is possible in a controlled manner or  whether KKLT can be controlled
in a strongly warped throat geometry.  In the latter respect, we pointed out that  
a destabilization of the uplifted dS minimum can occur (for $\sqrt{g_s} M < 6.8$) and that 
there exists a tower of highly red-shifted KK modes localized near the
tip of the throat that could potentially spoil the control over the
utilized low energy effective action. 

Assuming and being confident that  at least the AdS minimum (prior to uplift) \`a la KKLT
does exist, we reviewed how this relates to the ``classical'' AdS and dS swampland
conjectures,  motivating the introduction of quantum $\log$-factors at
various places.
Concretely, we have proposed corrections to the AdS/moduli scale separation conjecture
as well as the AdS distance conjecture.
Both these quantum generalizations lead to weaker bounds than the classical conjectures. 
In particular, a $\log$-type scale separation is allowed between the AdS
space  and the lightest modulus.

Viewing KKLT from the emergence perspective,  through a remarkable interplay of 
swampland conjectures we found that the natural species scale cut-off of KKLT coincides with the gaugino condensate scale 
of the implicitly assumed confining gauge theory 
$\Lambda_{\rm sp}\sim \Lambda_{\rm SQCD}$. 
Finally, as a new result we found that after including $\alpha'$
corrections to the K\"ahler potential the appearing dS saddle point
has a life-time that is even smaller than the bound from the dS
swampland conjecture. Hence, in this case the $\log$-corrections
correction do not weaken the bound towards saturating the bound from TCC.



\end{document}